\documentclass[10pt, article, twocolumn]{IEEEtran}
 \pdfoutput=1
\usepackage[pdftex]{graphicx}
\usepackage{enumerate}
\usepackage{caption}
\usepackage{subcaption}
\usepackage{amsthm}
\usepackage{amssymb}
\usepackage{dsfont,physics}
\usepackage{setspace}
\usepackage[linesnumbered,ruled,vlined]{algorithm2e}
\usepackage{comment}
\usepackage{color}
\usepackage{float}
\usepackage{gensymb}
\usepackage{comment}
\usepackage{xcolor}
\usepackage{cuted}
\usepackage{soul}

\theoremstyle{plain}

\newtheorem{definition}{Definition}
\newtheorem{assumption}{Assumption}

\theoremstyle{definition}

\theoremstyle{remark}

\usepackage{mathtools} \newcommand{\normm}[1]{\left\lVert#1\right\rVert}

\newcommand{\expect}{\operatorname{\mathbb{E}}\expectarg}
\DeclarePairedDelimiterX{\expectarg}[1]{[}{]}{%
  \ifnum\currentgrouptype=16 \else\begingroup\fi
  \activatebar#1
  \ifnum\currentgrouptype=16 \else\endgroup\fi
}
\newcommand{\prob}{\operatorname{\mathbb{P}}\probarg}
\DeclarePairedDelimiterX{\probarg}[1]{(}{)}{%
  \ifnum\currentgrouptype=16 \else\begingroup\fi
  \activatebar#1
  \ifnum\currentgrouptype=16 \else\endgroup\fi
}
\newcommand{\innermid}{\nonscript\;\delimsize\vert\nonscript\;}
\newcommand{\activatebar}{%
  \begingroup\lccode`\~=`\|
  \lowercase{\endgroup\let~}\innermid 
  \mathcode`|=\string"8000
}

\def\*#1{\mathbf{#1}}
\def\&#1{\mathcal{#1}}
\def\.#1{\boldsymbol#1}
\def\^#1{\hat{#1}}
\def\[#1{\left #1}
\def\]#1{\right #1}

\DeclareMathOperator*{\argmin}{arg\,min\ }
\DeclareMathOperator*{\argmax}{arg\,max\ }

\DeclareMathAlphabet\mathbfcal{OMS}{cmsy}{b}{n}

\begin{document}
\bstctlcite{BSTcontrol}

\title{Sequential Learning of CSI for MmWave Initial Alignment\\
}

\author{\IEEEauthorblockN{Nancy Ronquillo, Sung-En Chiu  and Tara Javidi}
\vspace{2mm}

\IEEEauthorblockA{Department of Electrical and Computer Engineering \\
University of California, San Diego \\
Email: \{nronquil,suchiu,tjavidi\}@ucsd.edu}
}

\maketitle

\begin{abstract}
MmWave communications aim to meet the demand for higher data rates by using highly directional beams with access to larger bandwidth. An inherent challenge is acquiring channel state information (CSI) necessary for mmWave transmission. We consider the problem of adaptive and sequential learning of the CSI during the mmWave initial alignment phase of communication. We focus on the single-user with a single dominant path scenario where the problem is equivalent to acquiring an optimal beamforming vector, where ideally, the resulting beams point in the direction of the angle of arrival with the desired resolution. We extend our prior by proposing two algorithms for adaptively and sequentially selecting beamforming vectors for learning of the CSI, and that formulate a Bayesian update to account for the time-varying fading model. Numerically, we analyze the outage probability and expected spectral efficiency of our proposed algorithms and demonstrate improvements over strategies that utilize a practical hierarchical codebook. 
\end{abstract}
\vspace{-5mm}
\section{Introduction}
Communications at mmWave bands are a promising approach to meeting the demand for increasingly higher data rates due to the availability of larger bandwidth at these frequencies. However, due to higher pathloss than traditional communications at sub 6~GHz with conventional antennas \cite{Maccartney2013}-\cite{Rappaport2017}, mmWave communications must be developed to overcome these challenges. A way to counter this problem is by increasing the signal gains at the transmitter and/or user ends. This can be achieved by leveraging the small wavelengths to fit many antennas in a small space that propagate highly directional beams \cite{Molisch2017}.  Under these highly directional beams, the challenge becomes learning the necessary channel state information (CSI).  We consider a single-path, and low-power single RF chain set-up where this initial alignment procedure is equivalent to acquiring directionality (i.e. learning the angle of arrival AoA). We are particularly concerned with enabling robustness to variations of the channel dynamics.

Initial alignment is well studied with many approaches implementing for example: Compressive Sensing (CS), Least squares (LS), Approximate Message Passing (AMP) or Matching Pursuit (MP) techniques, see
\cite{Abari2016, Song_TWC2018, Heath_Overview} 
and references therein. For a fixed power allocation and fixed very low overhead setting these approaches can handle alignment with no prior CSI despite a time-varying fading model with the caveat a sufficiently large SNR, or custom codebooks for improving SNR. This work follows most closely \cite{Alkhateeb2014}, where a fixed size hierarchical codebook and a bisection algorithm for sequentially selecting beamforming vectors to obtain the AoA are proposed.
Our prior work \cite{ChiuJSAC2019}, which proposes a fully adaptive initial alignment method based on posterior matching, theoretically characterizes an upper-bound on the probability of error in the AoA acquisition under a known static fading coefficient $\alpha$, and shows a significant improvement in the system communication rate over \cite{Alkhateeb2014} and random beamforming method of \cite{Abari2016}. For a slightly mismatched estimate of the fading coefficient $\alpha$ (i.e. a very good estimate) these improvements hold even under a time-varying channel. However, the performance is highly dependent on the quality of the channel knowledge. In fact, in the absence of a good estimate of $\alpha$ and under a time-varying fading model the performance is quite poor. 

In this study, we extend our prior work by proposing two methods that simultaneously and adaptively learn the fading coefficient as well as the AoA. The idea is to mitigate the effects of imperfect channel knowledge under a dynamic channel by augmenting the detection of the AoA to include simultaneous estimation of $\alpha$ for a given user. We propose two algorithms that work with the searching methods of our prior work \cite{ChiuJSAC2019} in order to fully learn the CSI (AoA and $\alpha$). Specifically, the contributions of the paper are as follows:  

\begin{itemize}
    \item First, we propose a direct extension of our prior work \cite{ChiuJSAC2019} to be robust to a time-varying fading model by adapting the posterior matching based strategy to include estimation of $\alpha$. 

    \item In light of excessive computations required to compute a Baysian posterior, we develop a heuristic approximation with low memory complexity. This  strategy compliments the sequential detection of AoA with estimation of the fading coefficient via a Kalman filter.

    \item 
    The proposed algorithms are compared to  our prior work \cite{ChiuJSAC2019} and the bisection algorithm of \cite{Alkhateeb2014} using the performance measures of outage probability and expected spectral efficiency. Numerically, we show improvements over \cite{Alkhateeb2014} and \cite{ChiuJSAC2019}, which suggests promising performance by our strategies for combining the learning of the AoA of the fading coefficients in the relevant regime of low ($-10$dB to $+5$dB) raw SNR. 
\end{itemize}
\underline{NOTATIONS:} We use boldface letters to represent vectors.
We denote the space of probability mass functions on set $\mathcal{X}$ as $P(x)$. 
$\mathcal{CN}(\.\mu, \.\Sigma)$ denotes the complex circularly symmetric Gaussian distribution. $\Re{c}, \Im{c}$ denote the real and imaginary parts of a complex number $c$, respectively. 
\section{Problem Set-up}

We consider a set-up with a Base Station (BS), equipped with $N$ antennas, serving $K$ users (UE), each with a fixed beamforming acting as single virtual antenna. We consider a low-power set-up where the BS and all users use a single RF Chain. We use a pilot-based procedure where the users send pilots to the BS while the BS combines the signal from the antenna elements to the RF chain by the beamforming vector $\mathbf{w_t}$ at time $t = 1,2, \ldots, \tau$, where each time represents a beamforming slot. 

For small-scale channel modelling, we use the stochastic multi-path modelling (see Ch.7 in \cite{Tse2005}) assumption with a single dominant path. 
\begin{assumption}\label{assum:singlepath}
The small-scale channel can be described as:
\begin{equation} \label{eq:singlepath}
    \*h = \alpha_t \*a(\phi),
\end{equation}
where $\alpha_t \in \mathbb{C}$ is the time-varying fading coefficient and 
\begin{equation}
    \*a(\phi) :=  [ 1, e^{j\frac{2\pi d }{\lambda} \sin{\phi} },..., e^{j(N-1)\frac{2\pi d }{\lambda} \sin{\phi} }  ]
\end{equation}
\label{singlepath_alpha}
is the array manifold created by the Angle-of-Arrival (AoA) $\phi\in[\theta_{\text{min}},\theta_{\text{max}}]$ with antenna spacing $d$. 
\end{assumption}

During the pilot procedure, we assume that the sequences corresponding to different users are orthogonal. i.e. for user $k \in K$ sending sequence $s_k$ we have the following:
\begin{assumption} \label{assum:cdma}
\begin{equation}
     \mathbf{s}_k^H \mathbf{s}_{k'} = \begin{cases} 1 \text{ for } k = k'\\
0 \text{ for }k \neq k'\end{cases}
\end{equation}
\end{assumption}
The code-matched signal from a particular user is given by
\begin{equation}\label{eq:obsv}
\begin{aligned}
    y_t &{=}  \sqrt{P} \*w_t^H  (\sum_{k'=1}^K \*h_{k'}  \mathbf{s}_{k'}^{T} ) \mathbf{s}_k^{*} + \*w_t^H \*N_{t} \mathbf{s}_k^*\\
        &\stackrel{(a)}{=}  \alpha_t \sqrt{P} \*w_t^H \*a(\phi) + \*w_t^H \*n_{t},
\end{aligned}
\end{equation}
where $\sqrt{P}$ is the combined transmit power and large scale fading, and the additive noise vector $\*n_{t} :=  \*N_{t} \mathbf{s}_k^* \sim \mathcal{CN}(\*0_{N\times 1},\sigma^2 \*I)$. We consider perfect knowledge of operating raw SNR, defined as $\frac{P}{\sigma^2}$ which is the SNR that would be received without beamforming. In this work we focus on relative performance, although the physical properties corresponding to this range of raw SNR (like cell size, and bandwidth) can be defined as in (Fig.~6 in \cite{ChiuJSAC2019}).
\vspace{-3mm}
\subsection{Sequential Initial Alignment}
\begin{figure}
    \centering
    \includegraphics[width = 0.3 \textwidth]{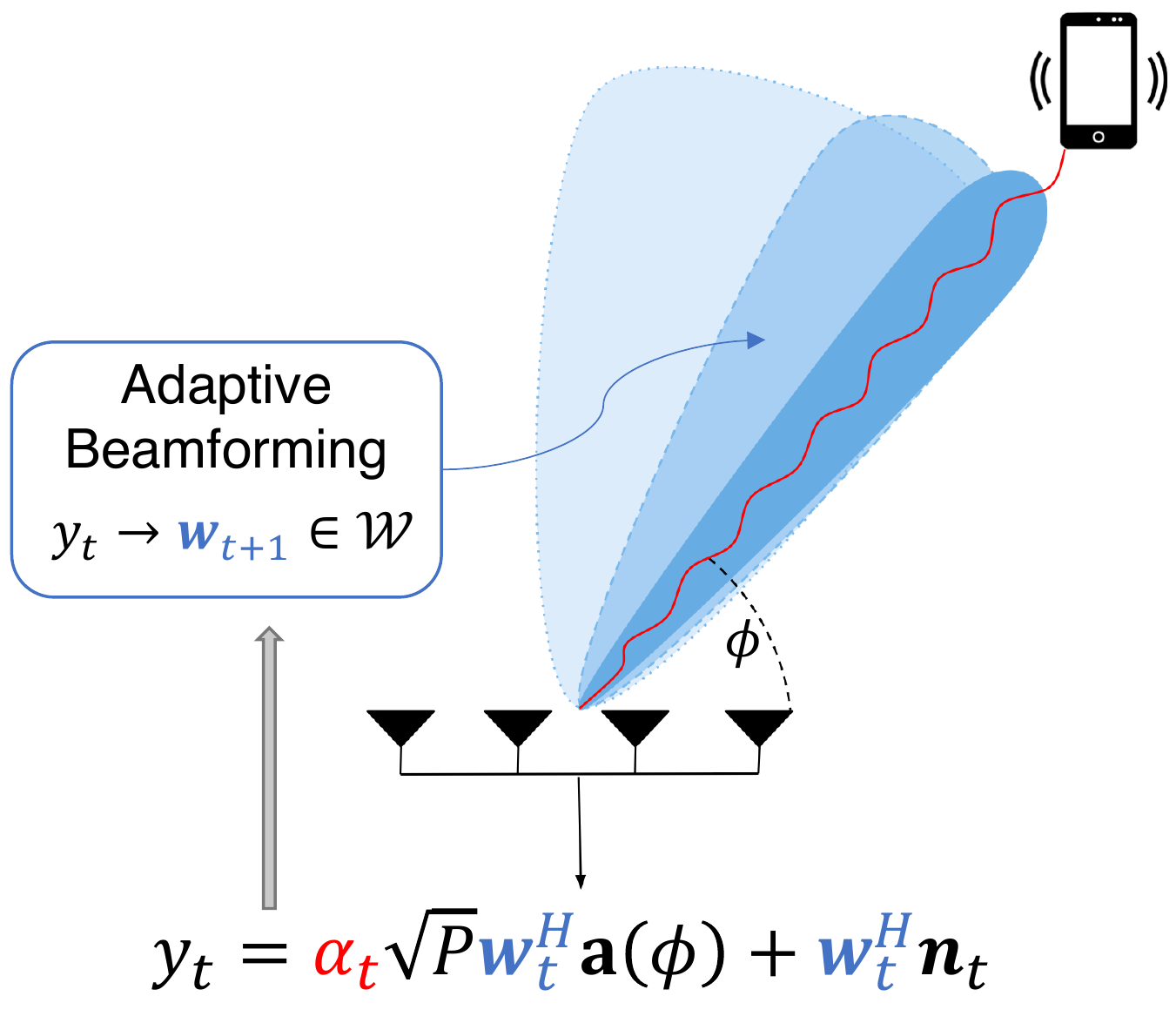}
    \caption{Active initial alignment is the process of sequentially selecting the beamforming vectors $\mathbf{w_t}(y_t)$, for a pilot-based procedure, based on prior observations $y_t$, after which the best directional beamforming vector for transmission is selected. }
    \label{fig:beamforming_yt}
\end{figure}
For the setup above, the BS must find the best directional beamforming vectors in order to establish a link to a given user, it does so via the process of initial alignment. The initial alignment procedure consists of selecting the best beamforming vector at each beamforming time slot $t = 1,2, \ldots, \tau$, possibly sequentially in an adaptive manner. After $\tau$ beamforming slots, a final beamforming vector $\hat{\mathbf{w}}$ is selected for subsequent transmission. In order to reduce reconfiguration time between beamforming slots, we utilize a codebook where:
\begin{assumption}
The beamforming vector is chosen from a pre-designed beamforming codebook $\mathcal{W}^S$ with finite cardinality. 
\end{assumption}


The quality of the established link, under assumption~\ref{eq:singlepath}, is determined by the accuracy of the final point estimate, $\hat{\phi}$, of $\phi$. In particular, a point estimate $\hat{\phi}$ together with a confidence interval $\delta$ provides robust beamforming with a certain outage probability. Hence, we measure the performance by the resolution and reliability of the final estimate $\hat{\*w}$:
\begin{definition}
Under Assumption~\ref{assum:singlepath}, a sequential beam search strategy  duration $\tau$, and final AoA estimate $\hat{\phi}$ is said to have resolution $\frac{1}{\delta}$ with error probability $\epsilon$ if
\begin{equation}
    \prob{|\hat{\phi}-\phi| > \delta} \leq \epsilon.
\end{equation}
\end{definition}



In this paper, we focus on strategies with sequential refinement of the beamforming vectors achieved by using the hierarchical beamforming codebook $\mathcal{W}^S$ of \cite{Alkhateeb2014} - described in Sect.~\ref{codebook}. More specifically, at any given time t, a beamforming vector $\mathbf{w_t}\in \mathcal{W}^S$ is selected sequentially as a function of previously observed signals $(y^{t-1})$. We focus on strategies based on posterior matching, where the problem is connected to channel coding over a binary input channel with (\cite{Giordani2016,Kaspi2018,Lalitha2017}) and without (\cite{Shabara2017}) feedback. We use the hierarchical posterior matching ($hiePM$) scheme from our prior work \cite{ChiuJSAC2019}, described in detail in Sect.~\ref{hiePMreview}, where the choice of $\mathbf{w_t}$ is such that the probability of $\phi$ lying in the area covered by $\mathbf{w_t}$ is closest to $\frac{1}{2}$. The effect is a sequential scanning of the angular space, where beamforming vectors become refined over time. 
\vspace{-3mm}
\subsection{Hierarchical Codebook}
\label{codebook}
The hierarchical codebook $\mathcal{W}^S$ consists of $S = log_2(1/\delta)$ levels that didactically partition the search space such that each level $l$ consists of a set $\mathcal{W}_l$ of $2^l$ possible beamforming vectors whose main beam width is $\frac{|\theta_{\text{max}}-\theta_{\text{min}}|}{2^l}$. Beamforming vectors have decreasing beam width as $l$ increases, resulting in  a codebook with a hierarchical structure\footnote[1]{Note that individual beamforming vectors $\mathbf{w}(\mathcal{D}_l^k) \in \mathcal{W}_l$ are designed with the objective of near constant gain for directions $\phi \in \mathcal{D}_l^k$ and almost zero otherwise. This is approximately achieved via a pseudo inverse approximation.}. The resulting beamforming weight vectors, applied with phase and gain control at each element, produce beam patterns with improved sidelobe suppression, and near constant gain in the intended search directions.   

\section{Jointly learning AoA and fading coefficient} 
We propose to use the sequential refinement adaptive strategy $hiePM$ from our prior work \cite{ChiuJSAC2019} in two proposed algorithms that extend $hiePM$ to simultaneously learn the CSI consisting of the AoA $\phi$, and the time-varying fading coefficient $\alpha_t$. First, we briefly review the $hiePM$ algorithm.

\subsection{Brief Review of hiePM}\label{hiePMreview}
\begin{figure}
    \centering
    \includegraphics[width = 0.4\textwidth]{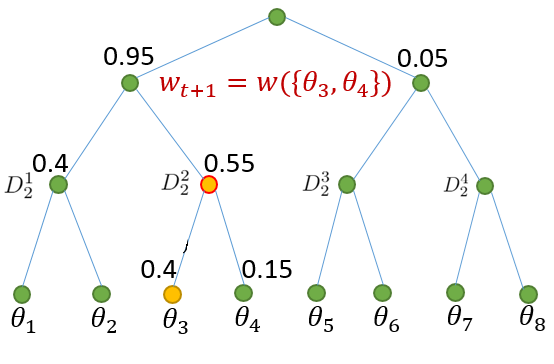}
    \caption{Illustration of $hiePM$ (Fig.~4 in \cite{ChiuJSAC2019}). In this example, we search down the tree hierarchy to levels 2 and 3, where level 3 has the first codeword that contains posterior lesser than half. Between level 2 and level 3, the codeword in level 2 of posterior $0.55$ is selected since it's closer to half ($0.55$ v.s. $0.4$). } 
    \label{fig:hiePM}
\end{figure}
$HiePM$ \cite{ChiuJSAC2019} is a method of sequentially selecting beamforming vectors $\*w_{t+1}$ from the codebook $\mathcal{W}^S$ based on the posterior probability vector for active learning of the AoA $\phi$. We resolve the AoA with a resolution $1 \slash\delta$ from a range of angles $[\theta_{\text{min}},\theta_{\text{max}}]$. For ease of exposition, we restrict the AoA to the discretized set $\phi \in \{ \theta_1,...,\theta_{1\slash \delta} \},\  \theta_i = \theta_{\text{min}} + (i-1)\times \delta \times (\theta_{\text{max}}-\theta_{\text{min}})$. The prior probability vector of $\phi$, is defined as $\.\pi^{\phi}(t)\in [0,1]^{1\slash \delta}$, where each element is: 
\begin{equation}
\label{eq:prior_phi}
    \pi_i(t) := \mathbb{P}{(\phi = \theta_i|\*y_{1:t})}, \ i=1,2,..., \frac{1}{\delta},
\end{equation}
and the probability of $\phi \in D_l^k$ (where $D_l^k$ is the range covered by the $k^{th}$ codeword in level $l$) can be computed as:
\begin{equation}
    \pi_{D_l^k}(t) := \sum_{\theta_i \in D_l^k} \pi_i(t).
\end{equation}
$HiePM$ selects $\*w_{t+1}(\boldsymbol{\pi}^{\phi}(t)) \in \mathcal{W}^S$ in a manner that sequentially refines the width of the beamforming vectors over time corresponding to the accumulated belief around the correct AoA $\phi$ as described by probability vector $\boldsymbol{\pi}^{\phi}(t)$, which is a sufficient statistic. $hiePM$ algorithm selects a codeword at either level $l^*$ or $l^*+1$ based on which codeword has probability closest to $\frac{1}{2}$. That is, $\*w_{t+1} (\boldsymbol{\pi}^{\phi}(t))$, is the $k_{t+1}^{th}$ codeword in level $l_{t+1}$ covering $D_{l_{t+1}}^{k_{t+1}}$ where:
\begin{equation}\label{hiePM_wt}
(l_{t+1},k_{t+1}) = \argmin_{(l',k')} \[|\pi_{D^{k'}_{l'}}(t) - \frac{1}{2}\]|    
\end{equation}
\vspace{-8mm}
\subsection{Computing $\.\pi^{\phi}(t)$}
In this section, we propose a Bayesian update as an extension to our prior work \cite{ChiuJSAC2019} which accounts for the random yet time-varying fading coefficient $\alpha_t$. Let us model the channel state at time $t$ as a random vector ($\phi, \underline{\alpha}_t) \in [\theta_\text{min}, \theta_\text{max}]\times \mathbb{C}$ with joint distribution $
f_{\phi,\underline{\alpha}_t}(\theta, \alpha),$
such that 
\begin{equation}
    \sum\limits_{i=1}^{1 \slash \delta}\int\displaylimits_{-\infty}^{\infty}\int\displaylimits_{-\infty}^{\infty} f_{\phi,\underline{\alpha}_t}(\theta_i, \alpha) d\Re{\alpha} d\Im{\alpha} =\sum\limits_{i=1}^{1 \slash \delta} \mathbb{P}(\phi=\theta_i)= 1. 
\end{equation}
The belief vector $\.\pi^{\phi}(t)$ can be computed sequentially:
\begin{equation}\label{phi_Bayes}
\begin{aligned}
    \mathbb{P}(\phi = \theta_i|y^{t})
        &=  \frac{    f_{Y_{t}|\phi,Y^{t-1}}(y_{t}| \theta_i, y^{t-1}) \mathbb{P}(\theta_i| y^{t-1})}
        {\sum\limits_{i' =1}^{1\slash \delta}f_{Y_{t}|\phi,Y^{t-1}}(y_{t}| \theta_i, y^{t-1}) \mathbb{P}(\theta_{i'}| y^{t-1})} \\
\end{aligned}
\end{equation}
where 
\begin{equation}
\begin{aligned}
& Y_{t}= \underline{\alpha}_t \sqrt{P} \*{w}_{t}^H \*{a}(\phi)+\underline{\eta}_t.\\
\end{aligned}
\end{equation}
where $(\underline{\eta}_t=\*{w}_{t}^H\*{n}_{t}) \sim \mathcal{CN}(0,\sigma^2)$\footnote[2]{Without loss of generality, the beamforming vectors in the codebook are assumed to have unit norm $\|\*w\|^2=1$.}. The conditional distribution is:
\begin{equation}\label{convolution}
\begin{aligned}
&f_{Y_{t}|\phi, Y^{t-1}}(y_{t}| \theta_i, y^{t-1}) \\
&	\approx  \int\displaylimits_{-\infty}^{\infty}\int\displaylimits_{-\infty}^{\infty} \frac{1}{G_i} f_{\underline{\alpha}_t|\phi,Y^{t-1}}\Big(\frac{y_{t}-\eta}{G_i}|\theta_i, y^{t-1}\Big) f_{\underline{\eta}_{t}}(\eta)d\eta\\
&\stackrel{(a)}{=}-\int\displaylimits_{-\infty}^{\infty}\int\displaylimits_{-\infty}^{\infty}  f_{\underline{\alpha}_t|\phi,Y^{t-1}}({\alpha}|\theta_i, y^{t-1}) f_{\underline{\eta}_{t}}(y_{t}-\alpha G_i)d{\alpha}\\
&\stackrel{(b)}{=}-\int\displaylimits_{-\infty}^{\infty}\int\displaylimits_{-\infty}^{\infty}  f_{\underline{\alpha}_t|\phi,Y^{t-1}}({\alpha}|\theta_i, y^{t-1}) g\Big(\frac{y_t-\alpha G_i}{\sigma}\Big) d{\alpha}\\
\end{aligned}
\end{equation}
where $d\eta =d\Re{\eta} d\Im{\eta}, d\alpha =d\Re{\alpha} d\Im{\alpha}$, 
\begin{equation}
    G_i = \sqrt{P} \mathbf{w_{t}}^H \mathbf{a}(\theta_i),
\end{equation}
 and (a) is by a change of variables, (b) follows by assumption of i.i.d noise, and where $g(\frac{y-\mu}{\sigma})$ is  $\mathcal{CN}(0, 1)$ evaluated at $(\frac{y-\mu}{\sigma})$. 
$ f_{\underline{\alpha}_t|\phi,Y^{t-1}}(\alpha|\theta_i, y^{t-1})$ can be obtained as a one-step prediction from  $f_{\underline{\alpha}_{t-1}|\phi,Y^{t-1}}(\alpha|\theta_i, y^{t-1})$. 
It follows that computation of $\boldsymbol{\pi}^{\phi}(t)$ depends on knowledge of 
$f_{\underline{\alpha}_{t}|\phi,Y^{t}}(\alpha|\theta_i, y^{t})$, a one-step prediction formulation for each time $t$, and may require infinite precision. We approach this computation by considering the following special cases. 

\subsubsection{Known and static fading coefficient $\underline{\alpha}_{t} = \alpha^*$:} For the case that the fading coefficient is static and known to the BS, we can write
\begin{equation}
    f_{\underline{\alpha}_t|\phi,Y^{t-1}}(\alpha| \theta_i, y^{t})=
    \begin{cases}
      1, & \text{if}\ \alpha=\alpha^* \\
      0, & \text{otherwise}.
    \end{cases}
\end{equation}
Plugging into Eq.~(\ref{phi_Bayes}) gives $\boldsymbol{\pi}^{\phi}(t)$ as:
\begin{equation}\label{eq:known_bayes}
\begin{aligned}
    \mathbb{P}(\phi=\theta_i| y^{t})
    &=\frac{ g\Big(\frac{y_t-\alpha^*G_i}{\sigma}\Big) 
    \mathbb{P}(\theta_i| y^{t-1})}{\sum\limits_{i' =1}^{1\slash \delta} g\Big(\frac{y_t-\alpha^*G_{i'}}{\sigma}\Big) 
    \mathbb{P}(\theta_i| y^{t-1})},
\end{aligned}
\end{equation}
which is finite, and does not require a one-step prediction of the conditional density of $\alpha$. This is recovers the formulation of our prior work \cite{ChiuJSAC2019}. 

\subsubsection{i.i.d. complex Gaussian fading coefficient with mean $\mu_{\tilde{\alpha}}$ and variance $\sigma^2_{\tilde{\alpha}}$.} In this case:
\begin{equation}
    f_{\underline{\alpha}_{t}|\phi,Y^{t-1}}(\alpha| \theta_i, y^{t-1})=f_{\underline{\alpha}_{t}}(\alpha)\sim \mathcal{CN}(\mu_{\tilde{\alpha}}, \sigma^2_{\tilde{\alpha}})
\end{equation}
is i.i.d for all $t$. Plugging into Eq.~(\ref{phi_Bayes}) gives the update $\boldsymbol{\pi}^{\phi}(t)$:
\begin{equation}\label{eq:iid_Bayes}
\begin{aligned}
    &\mathbb{P}(\phi = \theta_i| y^{t})\\
    &=\frac{\Big(\int\limits_{-\infty}^{\infty}\int\limits_{-\infty}^{\infty} g\Big(\frac{\alpha-\mu_{\tilde{\alpha}}}{ \sigma_{\tilde{\alpha}}}\Big) g\Big(\frac{y_t-\alpha G_i}{ \sigma}\Big) d{\alpha}  \Big)
     \mathbb{P}(\theta_i| y^{t-1})}{\sum\limits_{i' =1}^{1\slash \delta}\Big(\int\limits_{-\infty}^{\infty}\int\limits_{-\infty}^{\infty}  g\Big(\frac{\alpha-\mu_{\tilde{\alpha}}}{ \sigma_{\tilde{\alpha}}}\Big) g\Big(\frac{y_t-\alpha G_{i'}}{ \sigma}\Big) d{\alpha} \Big) \mathbb{P}(\theta_i| y^{t-1})}\\
    &=\frac{ g\Big(\frac{y_t-\mu_{\tilde{\alpha}} G_i}{ \sqrt{\sigma^2_{\tilde{\alpha}} \abs{G_i}^2 + \sigma^2}}\Big)  \mathbb{P}(\theta_i| y^{t-1})}{\sum\limits_{i' =1}^{1\slash \delta}   g\Big(\frac{y_t-\mu_{\tilde{\alpha}} G_{i'}}{ \sqrt{\sigma^2_{\tilde{\alpha}} \abs{G_{i'}}^2 + \sigma^2}}\Big)  \mathbb{P}(\theta_{i'}| y^{t-1})}.\\
\end{aligned}
\end{equation}
Note that E.q.~{(\ref{eq:iid_Bayes})} differs from the perfect knowledge scenario E.q.~{(\ref{eq:known_bayes})} only in $f_{Y_{t}|\phi, Y^{t-1}}(y_{t}| \theta_i, y^{t-1})$. E.q.~{(\ref{eq:iid_Bayes})} accounts for the uncertainty on the knowledge of the fading coefficient by increasing the variance by $(\sigma^2_{\tilde{\alpha}} \abs{G_i}^2)$, where the effect is a more conservative update. 

\subsubsection{Static fading coefficient $\underline{\alpha}_{t} = \underline{\alpha}$}\label{case3} Here we make the simplification that the fading coefficient is static in time \begin{equation}\label{pdf_static}
    f_{\underline{\alpha}_{t+1}|\phi,Y^{t}}(\alpha| \theta_i, y^{t})=f_{\underline{\alpha}|\phi,Y^{t}}(\alpha| \theta_i, y^{t}).
\end{equation}


Of course the Bayes' joint posterior probability update, detailed in Eq.~(\ref{phi_Bayes}), may be computed directly for a continuous density $f_{\underline{\alpha}|\phi,Y^{t}}(\alpha| \theta_i, y^{t})$, however, this may be computationally infeasible. In our first proposed Alg.~1, we approach the computation of $\boldsymbol{\pi}^{\phi}(t)$ by first discretizing $\underline{\alpha}$ over a finite number of sets.  To do this, let us make the simplification that $\underline{\alpha}= r_j+\text{i}z_k$ for $j = \{1, 2,\ldots, \frac{1}{\triangle_r}\}$ and $k = \{1, 2,\ldots, \frac{1}{\triangle_z}\}$ denotes that
\begin{equation}
\begin{aligned}
&\Re{\underline{\alpha}} \in [r_j-{\triangle_r \slash 2}, r_j+{\triangle_r \slash 2}]\\
&\Im{\underline{\alpha}} \in [z_k-{\triangle_z \slash 2}, z_k+{\triangle_z \slash 2}],
\end{aligned}
\end{equation}
where
\begin{equation}
\begin{aligned}
&r_j = r_{\text{min}} + (j-1)\times \triangle_r \times (r_{\text{max}}-r_{\text{min}})\\
&z_j = z_{\text{min}} + (z-1)\times \triangle_z \times (z_{\text{max}}-z_{\text{min}}),
\end{aligned}
\end{equation}
for $\Re{\underline{\alpha}} \in [r_{\text{min}},r_{\text{max}}]$, and $\Im{\underline{\alpha}} \in [z_{\text{min}},z_{\text{max}}]$. Under this discretization, the probability is:
\begin{equation}\label{discrete_probalpha}
    \mathbb{P}(\underline{\alpha} = r_j+\text{i}z_k|\theta_i y^{t}) =
    f_{\underline{\alpha}}(r_j+\text{i}z_k)\triangle_r\triangle_z
\end{equation}
We note that this discretization becomes tight as $\triangle_r,\triangle_z \rightarrow 0$. Coarser choices of ($\triangle_r$, $\triangle_z$) reduce complexity and memory requirements. (see Sect.~\ref{sims} for a discussion of practical values for these resolution parameters used in our simulations).

In this paper, we propose two strategies developed for this case of a static fading. In the first, given in Alg.~1, beamforming vectors are sequentially selected according to $hiePM$, where the posterior belief on $\phi$, $\boldsymbol{\pi}^{\phi}(t)$, is used to make the beam selection from $\mathcal{W}^S$. 
For a received observation $y_{t}$ each element of the posterior update on the belief on $\phi$, $\boldsymbol{\pi}^{\phi}(t)$, can be obtained by Eq.~(\ref{phi_Bayes}) and ~(\ref{discrete_probalpha}):
\begin{equation}\label{eq:JPM_discrete}
\begin{aligned}
    &\mathbb{P}(\phi =\theta_i| y^{t})=\\
    &\frac{\sum\limits_{j=1}^{{1}\slash{\triangle_r}}\sum\limits_{k=1}^{{1}\slash{\triangle_z}}g\Big(\frac{y_t-(r_{j}+\text{i}z_{k})G_{i}}{\sigma}\Big)\mathbb{P}{(r_j,z_k|\theta_i, y^{t-1})}\mathbb{P}{(\theta_i| y^{t-1})}}
    {\sum\limits_{j'=1}^{1 \slash \triangle_r}\sum\limits_{k'=1}^{1 \slash \triangle_z}\sum\limits_{i'=1}^{{1}\slash{\delta}}  g\Big(\frac{y_t-(r_{j}+\text{i}z_{k})G_{i'}}{\sigma}\Big)  \mathbb{P}{(r_{j'},z_{k'}|\theta_i, y^{t-1})}\mathbb{P}{(\theta_{i'}| y^{t-1})}} \\
\end{aligned}
\end{equation}
Note, we are required to keep track of the joint probability $\boldsymbol{\pi}^{\phi,\underline{\alpha}}(t)\in [0,1]^{\frac{1}{\delta} \times \frac{1}{\triangle_r} \times \frac{1}{\triangle_z}}$ at all times $t$, defined as:
\begin{equation}
\begin{aligned}
\pi_{i,j,k}(t) = \mathbb{P}{(r_j,z_k|\theta_i, y^{t})}\mathbb{P}{(\theta_i| y^{t})}.
\end{aligned}
\end{equation}
Alg.~1 rewrites Eq.~(\ref{eq:JPM_discrete}) in two steps first with an update to get $\boldsymbol{\pi}^{\phi,\underline{\alpha}}(t)$, and then a marginalization to get $\.\pi^{\phi}(t)$ and thus has computational cost on the order of $O(\frac{1}{\triangle_r} \times \frac{1}{\triangle_z} \times \log(\frac{1}{\delta}))$.
\vspace{-5mm}
\begin{algorithm}[h!tb] 
 \textbf{Input}: target resolution $({\delta},\triangle_r, \triangle_z)$, codebook $\mathcal{W}^S$ $(S = \log_2(1\slash \delta))$, $\tau$  (length of the initial access phase)\\
 \textbf{Output}: Estimate of the AoA $\hat{\phi} $\\
 \textbf{Initialization}: $\boldsymbol{\pi}^{\phi,\underline{\alpha}}(0):\pi_{i,j,k}(0) = \delta\triangle_r \triangle_z$  $\forall i, j, k$ \\
 \For{$t=1,2,..., \tau$ }
 {
     \# Marginal posterior of $\phi$, $\boldsymbol{\pi}^{\phi}(t)$:
     \begin{gather}
        \label{eq:marginal_phi}
         \pi_{i}(t+1) = \sum\limits_{j=1}^{{1}\slash{\triangle_r}}\sum\limits_{k=1}^{{1}\slash{\triangle_z}} \pi_{i,j,k}(t+1).
     \end{gather}\\
    \# Beam selection according to $hiePM$ Eq.~(\ref{hiePM_wt}):
     \begin{gather*}
         \*w_{t+1}(\boldsymbol{\pi}^{\phi}(t)) \in \mathcal{W}^S
     \end{gather*}\\
    \# Take next measurement \\[-4mm]
     \begin{gather*} 
     \begin{aligned} \label{eq:measure}
         y_{t+1} &= \underline{\alpha}_{t+1} \sqrt{P} \*w_{t+1}^H \*a(\phi) + \*w_{t+1}^H \*n_{t+1} \\
     \end{aligned}
     \end{gather*}\\
    \# Joint posterior update by Bayes' Rule Eq.~(\ref{eq:JPM_discrete}) \\[-4mm]
     \begin{gather*} 
     \label{eq:Bay}
         \.\pi^{\phi,\underline{\alpha}}(t+1) \leftarrow y_{t+1}, \.\pi^{\phi,\underline{\alpha}}(t) 
     \end{gather*} \\
 }
     \# Final beamforming vector design \\[-4mm]
         \begin{gather}\label{eq:Fres}
         \hat{\phi} = \argmax_{\theta_i} \pi_i(\tau)
     \end{gather}\\[-4mm]
   $\hat{\*w} = \*w(\hat{\phi})$
\caption{Joint Posterior Matching}
\end{algorithm}

We also propose an alternative method, detailed in Alg.~2, which reduces the computational cost to $O(\frac{1}{\delta} \times \log(\frac{1}{\delta}))$ by implementing the Kalman filter \cite{Fundamentals_SSP}. First, we assume the conditional probability density Eq.(\ref{pdf_static})  is complex Gaussian: 
\begin{equation}
          f_{\underline{\alpha}|\phi,Y^{t}}(\alpha| \theta_i, y^{t})\sim \mathcal{CN}(\mu_{\alpha,i}(t) , \sigma^2_{\alpha,i}(t)), 
\end{equation}
with known prior $(\mu_{\underline{\alpha},i}(0), \sigma^2_{\underline{\alpha},i}(0))$ for all $i$.

The proposed algorithm (summarized in Alg.~2) selects beamforming vectors sequentially according to $hiePM$,  where a  marginalized probability over $\phi$, $\boldsymbol{\pi}^{\phi}(t)$, is used to make the beam selection from $\mathcal{W}^S$. Upon receiving a new observation $y_{t+1}$, the mean and variance of this conditional probability are updated by the Kalman filter:
\begin{equation}
\label{eq:Kalman}
\begin{aligned}
\mu_{\underline{\alpha},i}(t+1)&= \mu_{\underline{\alpha},i}(t) + \frac{ \sigma^2_{\underline{\alpha},i}(t) \overline{G_i} }{ \sigma^2_{\underline{\alpha},i}(t)\abs{G_i}^2 + \sigma^2 } (y_{t+1}-\mu_{\underline{\alpha},i}(t)G_i )\\
 \sigma^2_{\underline{\alpha},i}(t+1) &= \sigma^2_{\underline{\alpha},i}(t) \frac{ \sigma^2 }{ \sigma^2_{\underline{\alpha},i}(t)\abs{G_i}^2 + \sigma^2 }.
\end{aligned}
\end{equation}
Next, Alg.~2 uses the estimate of the fading coefficient to obtain  $\boldsymbol{\pi}^{\phi}(t)$ by Eq.~(\ref{phi_Bayes}) and~(\ref{eq:Kalman}): 
\begin{equation}\label{eq:Kalman_Bayes}
\begin{aligned}
    &\mathbb{P}(\theta_i| y^{t})=\\
    &\frac{\Big(\int\limits_{-\infty}^{\infty}\int\limits_{-\infty}^{\infty} g\Big(\frac{\alpha-\mu_{\underline{\alpha},i}(t) }{ \sigma_{\underline{\alpha},i}(t)}\Big) g\Big(\frac{y_t-\alpha G_i}{ \sigma}\Big)\Big)d{\alpha} \mathbb{P}(\theta_i| y^{t-1})}{\sum\limits_{i' =1}^{1\slash \delta}\Big(\int\limits_{-\infty}^{\infty}\int\limits_{-\infty}^{\infty}  g\Big(\frac{\alpha-\mu_{\underline{\alpha},{i'}}(t) }{ \sigma_{\underline{\alpha},{i'}}(t)}\Big) g\Big(\frac{y_t-\alpha G_{i'}}{ \sigma}\Big) \Big)d{\alpha} \mathbb{P}(\theta_i| y^{t-1})}\\
    &=\frac{g\Big(\frac{y_t-\mu_{\underline{\alpha},i}(t) G_i}{\sqrt{\sigma^2_{\underline{\alpha},i}(t) \abs{G_i}^2 + \sigma^2}}\Big) \mathbb{P}(\theta_i| y^{t-1})}{\sum\limits_{i' =1}^{1\slash \delta} g\Big(\frac{y_t-\mu_{\underline{\alpha},{i'}}(t) G_{i'}}{\sqrt{\sigma^2_{\underline{\alpha},{i'}}(t) \abs{G_{i'}}^2 + \sigma^2}}\Big)  \mathbb{P}(\theta_i| y^{t-1})}.\\
\end{aligned}
\end{equation}
\vspace{-5mm}
\begin{algorithm}[h!tb] 
 \textbf{Input}: target resolution ${\delta}$, codebook $\mathcal{W}^S$ $(S = \log_2(1\slash \delta))$, $\tau$  (length of the initial access phase)\\
 \textbf{Output}: Estimate of the AoA $\hat{\phi} $\\
 \textbf{Initialization}:$\boldsymbol{\pi}^{\phi}(0):\pi_{i}(0) = \delta \ \forall i,$  $(\boldsymbol{\mu}_{\underline{\alpha}}(0), \boldsymbol{\sigma}^2_{\underline{\alpha}}(0))$ \\
 \For{$t=1,2,..., \tau$ }
 {
    \# Beam selection according to $hiePM$ Eq.~(\ref{hiePM_wt}):
     \begin{gather*}
         \*w_{t+1}(\boldsymbol{\pi}^{\phi}(t)) \in \mathcal{W}^S
     \end{gather*}\\
    \# Take next measurement \\[-4mm]
     \begin{gather*} 
     \begin{aligned} \label{eq:measure_
    KL}
         y_{t+1} &= \underline{\alpha}_{t+1} \sqrt{P} \*w_{t+1}^H \*a(\phi) + \*w_{t+1}^H \*n_{t+1} \\
     \end{aligned}
     \end{gather*}\\
    \# Update moments by Kalman Filter Eq.~(\ref{eq:Kalman}) \\[-4mm]
     \begin{gather*} 
     \label{eq:Bay}
         (\.\mu_{\underline{\alpha}}(t+1), \.\sigma^2_{\underline{\alpha}}(t+1))\leftarrow y_{t+1}, \.\mu_{\underline{\alpha}}(t), \.\sigma^2_{\underline{\alpha}}(t)
     \end{gather*} \\
    \# Posterior update by Bayes' Rule Eq.~(\ref{eq:Kalman_Bayes}) \\[-4mm]
     \begin{gather*} 
     \label{eq:Bay}
         \.\pi^{\phi}(t+1) \leftarrow y_{t+1}, \.\pi^{\phi}(t), \.\mu_{\underline{\alpha}}(t), \.\sigma^2_{\underline{\alpha}}(t)  
     \end{gather*} \\
 }
     \# Final beamforming vector design 
     \\[-4mm]
         \begin{gather}\label{eq:Fres}
         \hat{\phi} = \argmax_{\theta_i} \pi_i(\tau)
     \end{gather}\\[-4mm]
   $\hat{\*w} = \*w(\hat{\phi})$
\caption{Kalman Filter for Posterior Matching}
\end{algorithm}
\vspace{-8mm}
\section{Numerical Results}
\label{sims}
Our numerical simulations analyze the proposed algorithms using the performance measures of outage probability and achieved spectral efficiency in the the relevant regime of low (-$10$dB to +5dB) raw SNR. 
\vspace{-3mm}
\subsection{Simulation Scenario}
We consider a scenario where the BS is equipped with a uniform linear array with $N = 64$ antenna elements with spacing $\frac{\lambda}{2}$. We focus on the single user case, where the UE has fixed beamforming acting as a single virtual antenna. We aim to learn the AoA with  resolution $1 \slash \delta = 128$.

Even though Alg.~1 and Alg.~2 are developed for the case of static fading Sect.~\ref{case3}, we apply these over a more practical AR-1 time correlated model, described below. Under perfect knowledge of the operating SNR (large-scale fading) as well as perfect frequency/phase synchronization, the fading coefficient is given as:
\begin{equation*}\label{AR-1}
    \alpha_{t+1} = \alpha_t \sqrt{1-g}  + \sqrt{\frac{k_r}{1+k_r}} \gamma \Bigg( 1- \sqrt{1-g}  \Bigg) + e_t\sqrt{\frac{g}{1+k_r}}, 
\end{equation*}
where $\gamma = 1$, $k_r$ is the Rician fading factor, $g$ is the correlation parameter, and $e_t\sim \mathcal{CN}(0,1)$ is the independent noise term. 

The correlation parameter $g$ is set such that 
coherence time $T_c=2$ ms (equivalent to $\tau =28$ total slots using the 5G NR PRACH format B4 \cite{Lin20185GNR}), and a Rician factor $k_r=10$ (this is a reasonable value, e.g. indoor mmWave channel models \cite{Kfactor_MUKHERJEE}). 
For these parameters, we assume a conservative range for $\underline{\alpha}_t$ to be: $[r_{\text{min}},r_{\text{max}}] = [0,2], \ [z_{\text{min}},z_{\text{max}}] = [-0.7, 0.7]$,
with resolution parameters $\frac{1}{\triangle_r} = \frac{1}{\triangle_z} = 50$ \footnote[3]{Very low resolution parameters will reduce complexity but may impact performance, a further investigation is outside the scope of this paper.}.
\vspace{-3mm}
\subsection{Probability of error}
\begin{figure}
    \centering
    \includegraphics[width = 0.45 \textwidth]{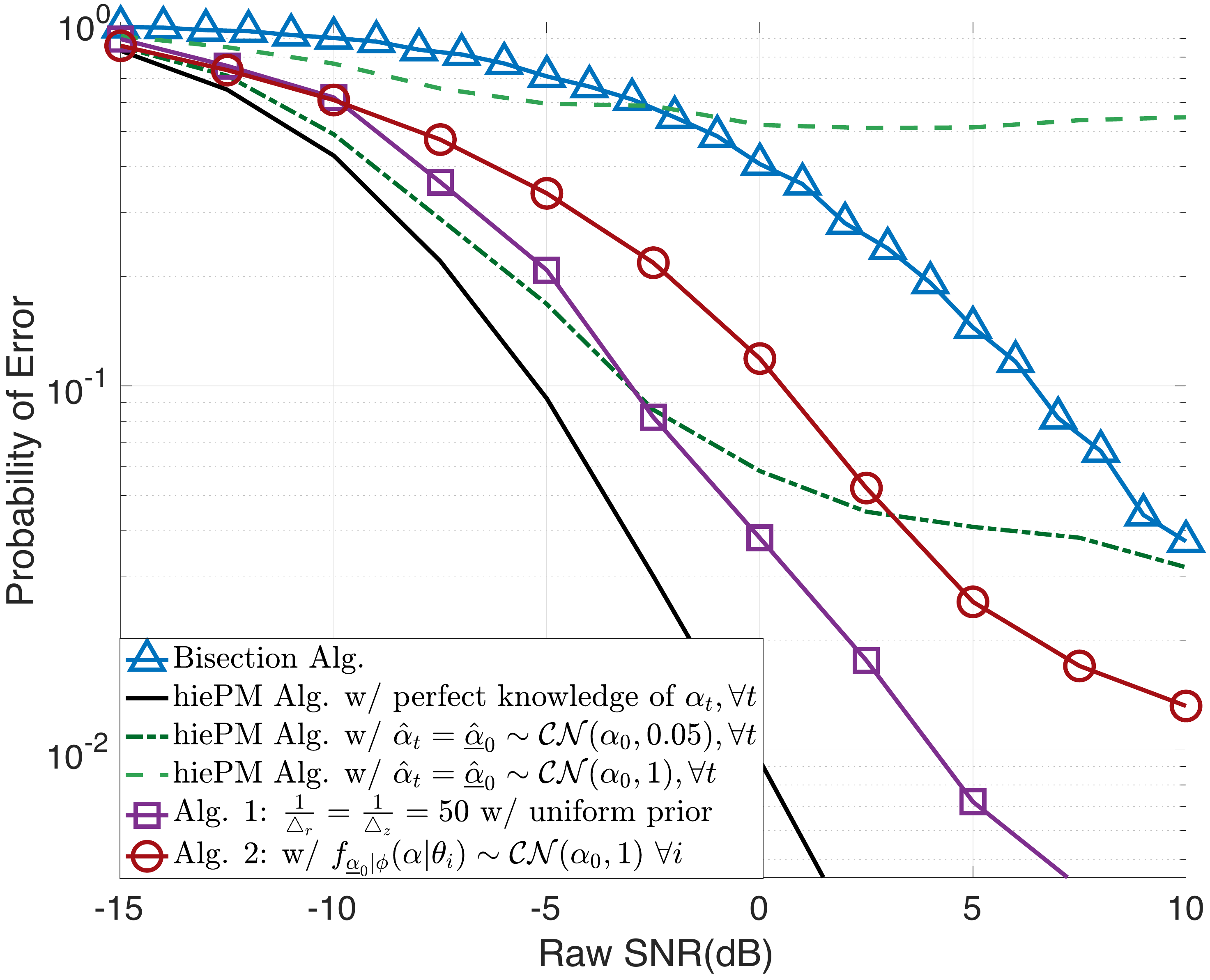}
    \caption{Comparison of the error probability performance between our proposed algorithms and prior works as a function of raw SNR $P/\sigma^2$ under Rician AR-1 fading. The probability of error in selecting the correct final beamforming is given by Eq.~(\ref{eq:poe}). }   
    \label{PoE_graph}
\end{figure}
First, we analyze the performance of our proposed algorithms in terms of the probability of error in choosing the correct final beamforming vector, defined as:
\begin{equation}\label{eq:poe}
\text{Prob}\{{\textbf{w}}(\hat{\phi}) \neq \textbf{w}(\phi)\}.
\end{equation}
In Fig.~\ref{PoE_graph} we compare performance to the Bisection algorithm of \cite{Alkhateeb2014}, which utilizes no prior knowledge of of the fading coefficient, and to our prior work $hiePM$ of \cite{ChiuJSAC2019} which utilizes the mismatched guess $\hat{{\alpha}}=({\underline{\alpha}}_0\sim\mathcal{CN}(\alpha_0,\sigma^2_{\alpha}))$ for all $t$. First, we note that while $hiePM$ with mismatch can be robust to the time-varying fading, this performance is highly dependent on the quality of the estimate. 
Our proposed Alg.~1 closely approaches the performance of perfect knowledge $\hat{\alpha}_t=\alpha_t$ and outperforms \cite{ChiuJSAC2019} and \cite{Alkhateeb2014}. This suggests that Alg.~1 is in fact learning $\underline{\alpha}$ while simultaneously beamforming to detect $\phi$ in the duration $\tau$. Similarly, Alg.~2 also improves on \cite{ChiuJSAC2019} and \cite{Alkhateeb2014}. However, as expected the performance is not a good as Alg.~1 due to the assumption we make about shape of the estimate {$f_{\underline{\alpha}|\phi,Y^{t}}(\alpha| \theta_i, y^{t})$} in order to reduce the computation complexity. As a result, the proposed algorithms highlight a trade-off between computational complexity and performance in terms of probability of error. 
\vspace{-3mm}
\subsection{Spectral Efficiency}
\begin{figure}
    \centering
    \includegraphics[width = 0.45 \textwidth]{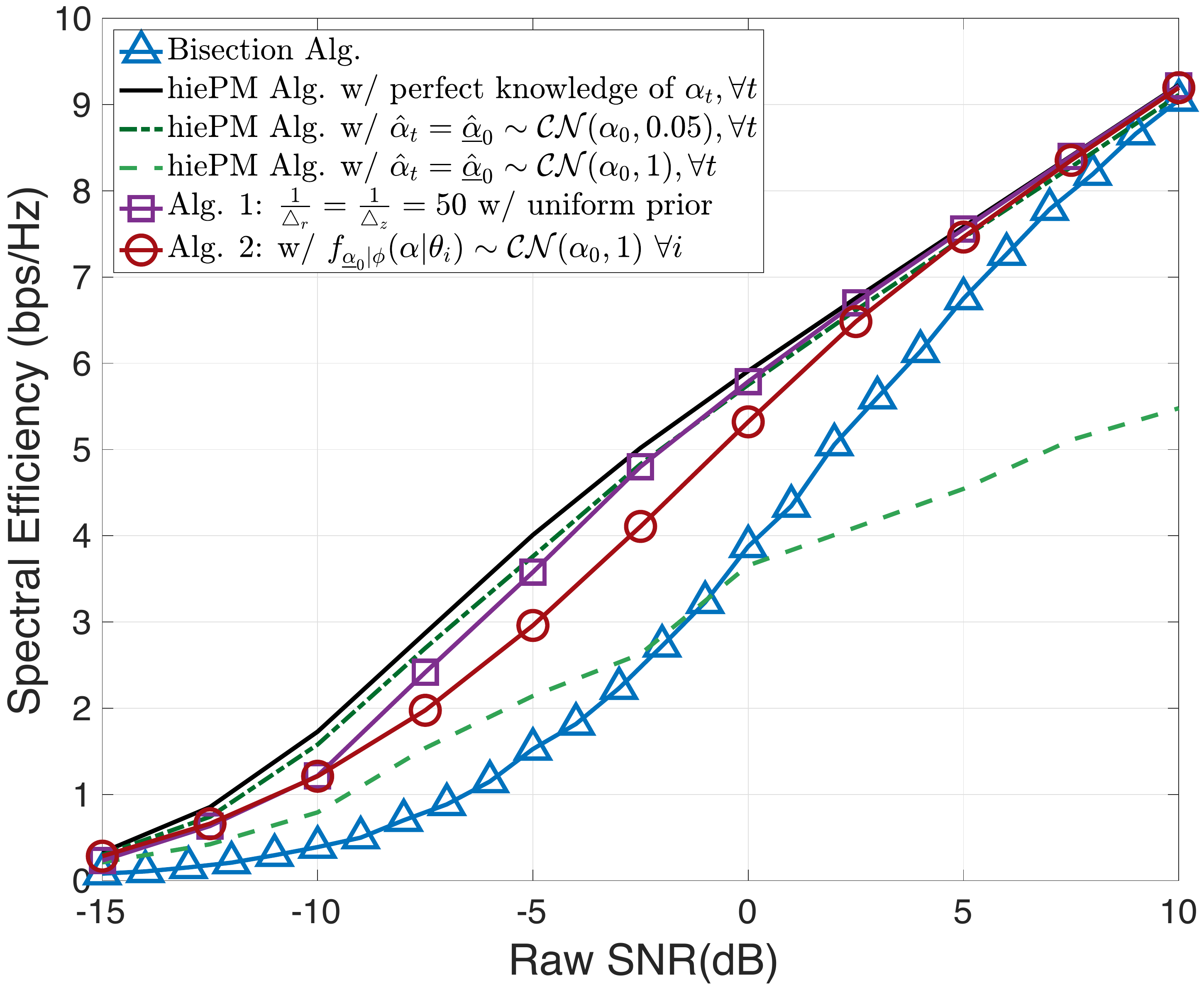}
    \caption{Comparison of the spectral efficiency between our proposed algorithms and prior works as a function of raw SNR $P/\sigma^2$ under Rician AR-1 fading. The spectral efficiency is given by Eq.~(\ref{eq:data_rate}).  }
    \label{RATE_graph}
\end{figure}

Next, we analyze the practical metric of spectral efficiency. Given a total communication time frame $T$\footnote[4]{$\tau$ and $T$, may require further system optimization, however, this is outside the scope of this paper.}, the expected spectral efficiency achieved using the final beamforming vector ${\*w}(\hat{\phi})$ is:
\begin{equation} \label{eq:data_rate}
    \expect*{\frac{T-\tau}{T} \log \[( 1+  \frac{P \mid {\*w}(\hat{\phi})^H \*a(\phi) \mid^2 }{\sigma^2}  \]) }.
\end{equation}
In Fig.~\ref{RATE_graph} we plot the performance in spectral efficiency Eq.~(\ref{eq:data_rate}) of our proposed algorithms compared to $hiePM$ with mismatch of \cite{ChiuJSAC2019} and the Bisection algorithm of \cite{Alkhateeb2014}. The results mimic closely the results we observed in performance of probability of error. 
Our proposed Alg.~1 and Alg.~2 achieve improvements spectral efficiency over \cite{ChiuJSAC2019} and \cite{Alkhateeb2014} due to the mistakes that these make in the selection of the correct final beamforming vector. We see how critical these mistakes are in the regime low ($-10$dB to $+5$dB) raw SNR where correct selection of the beamforming vector enables much higher spectral efficiency.
\section{Conclusion and Future Directions}
This work considers the problem of sequential beamforming design for MmWave initial alignment. We propose two algorithms that extend our prior work to learn the CSI made up of the fading coefficient and AoA. 
As a first step to demonstrate robustness under a practical time-varying fading model, our numerical results show that under a practical hierarchical codebook the proposed algorithms outperform our prior work, $hiePM$ with mismatched estimate of $\alpha$ and a strategy for sequential beam refinement in the literature (the Bisection algorithm of \cite{Alkhateeb2014}). Our ongoing work considers development of a kalman prediction for complexity reduction in Alg.~2 and a prediction formulation in the posterior for Alg.~1 for even better performance under a dynamic fading. Comparison to strategies that utilize larger, user specific, or carefully designed codebooks is left as a future work. Analysis under other types of fading with more complicated models like multi-path fading, and mobile users is also of practical interest. 
\bibliographystyle{IEEEtran}
\vspace{-3mm}
\bibliography{./refs}

\end{document}